\documentclass[aps,prb,reprint]{revtex4-2}

\usepackage{graphicx}
\usepackage{dcolumn}
\usepackage{bm}
\usepackage{amsmath}
\usepackage{amssymb}
\usepackage{hyperref}
\usepackage{float}
\usepackage{multirow}
\usepackage{mathrsfs}
\usepackage[title]{appendix}
\usepackage{xcolor}
\usepackage{comment}
\usepackage{mathtools}
\usepackage{anyfontsize}
\usepackage{textcomp}
\usepackage[caption=false]{subfig}
\usepackage[nameinlink,noabbrev]{cleveref}

\begin{document}

\title{Quasi-bound States of Scalar field inside the Dyonic Kerr-Sen Black Hole}

\author{David Senjaya}%
 \email{davidsenjaya@protonmail.com}
\affiliation{Department of Physics, Faculty of Science, Mahidol University, 272 Rama VI Road, Ratchathewi, Bangkok 10400, Thailand}

\author{Tinnagrit Songkeaw}
\email{sonprateep@gmail.com}

\author{Piyabut Burikham}
 \email{piyabut@gmail.com}
\affiliation{High Energy Physics Theory Group, Department of Physics, Faculty of Science,
Chulalongkorn University, Bangkok 10330, Thailand}

\date{\today}

\begin{abstract}

We found sets of exact analytic quasi-stationary states of a massive scalar field in a dyonic Kerr-Sen black hole~(DKSBH) background in the maximally extended spacetime region. A central novelty is the use of horizon-regular ingoing Eddington-Finkelstein coordinates, which enables a direct and unambiguous imposition of the ingoing boundary condition at the horizon. The exact radial solutions are in the form of confluent Heun functions. Imposing regularity at spatial infinity enforces a series truncation condition, yielding an exact quantization of the quasi-stationary frequencies. The spectrum exhibits a rich multi-branch structure, which we show splits into two distinct classes: modes that are insensitive to the black hole spin and charges and modes that explicitly depend on them. We uncover a clear asymmetry between co-rotating and counter-rotating configurations, driven by the spin-angular momentum coupling, as well as a systematic shift of the spectrum induced by electric and magnetic charges. The physical branches exhibit a universal behavior: modes with positive real frequency possess positive imaginary parts and therefore grow exponentially in time, whereas modes with negative real frequency are damped and decay. This suggests that positive-energy excitations in the region behind the outer horizon including the inner region of the inner horizon which contains the closed-timelike-curve, exponentially destabilize the background spacetime, supporting Hawking's chronology protection conjecture. In addition, the purely imaginary modes contain no oscillatory component and hence do not propagate through the spacetime, preventing traveling excitations along closed timelike curves and remaining consistent with the conjecture.

\end{abstract}




\maketitle

\section{Introduction}
The theory of general relativity, proposed by Einstein, stands as one of the most profound achievements in modern physics, successfully superseding the Newtonian description of gravity. In this geometric framework, gravity is not interpreted as a force but as a manifestation of spacetime curvature generated by mass and energy. The motion of particles and light is governed by geodesic equations determined entirely by the spacetime metric. Over the past century, numerous predictions of general relativity—including the perihelion precession of Mercury, gravitational lensing, gravitational redshift, gravitational waves, and the existence of black holes—have been confirmed experimentally with remarkable precision \cite{Hobson,Will,Abbott,Abbott1,EHTC}.

Despite these successes, general relativity is not without conceptual and phenomenological limitations. In particular, many black hole and cosmological solutions contain spacetime singularities, where curvature invariants diverge and the classical description ceases to be valid. Furthermore, the theory cannot account for the observed flat rotation curves of galaxies or the accelerated expansion of the universe without introducing dark components through the energy-momentum tensor $T_{\mu\nu}$. These issues have motivated the development of various extended theories of gravity, including $f(R)$ and $f(T)$ theories, Gauss-Bonnet gravity, scalar-tensor-vector gravity (STVG), Lovelock gravity, and Kaluza-Klein theory, among others \cite{Moffat_2006,MOFFAT_2007,Clifton_2012,Papantonopoulos:2015cva,Petrov_2020}.

In this work, we focus on the string-inspired Einstein-Maxwell-dilaton-axion (EMDA) theory, which belongs to the scalar-vector-tensor class of supergravity models. The EMDA framework extends Einstein-Maxwell theory by incorporating nontrivial couplings between the electromagnetic field tensor $F_{\mu\nu}$, the scalar dilaton field $\xi$, and the pseudo-scalar axion field $\phi$. The corresponding four-dimensional effective action is given by \cite{Sen_1992,Banerjee_2020}
\begin{multline}
S_{EMDA}=\frac{1}{16\pi}\int\left[R-2\partial_\mu\xi\partial^\mu\xi
-\frac{1}{3}H_{\rho\sigma\delta}H^{\rho\sigma\delta}
\right. \\ \left.+e^{-2\xi}F_{\alpha\beta}F^{\alpha\beta}\right]\sqrt{-g}\, d^4x,
\end{multline}
where $R$ denotes the Ricci scalar and $g$ is the determinant of the metric tensor. The Maxwell field tensor is defined in terms of the $U(1)$ gauge field $A_\mu$ as
\begin{equation}
F_{\mu\nu}=\partial_\mu A_\nu-\partial_\nu A_\mu,
\end{equation}
while the Kalb-Ramond field tensor is related to the pseudo-scalar axion field through
\begin{equation}
H_{\alpha\beta\delta}=\frac{1}{2}e^{4\xi}\varepsilon_{\alpha\beta\delta\gamma}\partial^\gamma \phi.
\end{equation}

In \cite{Wu_2021}, the authors construct an exact rotating black hole solution in EMDA gravity characterized by four independent charges: electric, magnetic, dilaton, and axion. This geometry, known as the dyonic Kerr-Sen black hole, generalizes the original Kerr-Sen solution \cite{Sen_1992} and reduces to the Kerr spacetime of Einstein-Maxwell theory~\cite{Kerr:1963ud} when the dilaton and axion fields are switched off. The presence of these additional charges enriches the horizon structure and modifies the effective potential governing scalar perturbations, making this spacetime a natural setting to explore quasi-stationary spectra beyond the standard Kerr-Newman case. From the perspective of wave dynamics, the main challenge is solving the Klein-Gordon equation in such a rotating and highly nontrivial background. In general, the resulting radial equation does not admit closed-form solutions, and one must rely on approximation methods, which often obscure the underlying analytic structure of the spectrum.

The novelty of this work is twofold: First, we show that the Klein-Gordon equation in the dyonic Kerr-Sen spacetime remains separable even in ingoing Eddington-Finkelstein coordinates, where separability is not a priori guaranteed. This allows the scalar dynamics to be reduced to a system of ordinary differential equations, with the radial part governed by a confluent Heun equation. Second, we derive an exact analytic solution of the radial equation without resorting to asymptotic or perturbative approximations. The solution is expressed in terms of confluent Heun functions, and imposing the polynomial (termination) condition leads to a discrete set of complex frequencies. These correspond to quasi-stationary states, whose real part determines the oscillation frequency while the imaginary part characterizes damping or growth.

In Ref.~\cite{Senjaya:2024gpb} and \cite{Bunyaratavej:2024qgk}, the exact solutions of the scalar field in the DKSBH in the outer~(inner) region of the {\it outer~(inner)} horizon are obtained by using the Boyer-Lindquist coordinates respectively. Since the metric of spacetime region between the inner and outer horizon changes signature as a direct consequence of the Boyer-Lindquist coordinates, the quasi-resonance modes in such region are not yet rigorously determined. In this work, the use of ingoing Eddington-Finkelstein coordinates ensure a regular and physically transparent description across the event horizon, avoiding the signature change of the metric and the coordinate singularities. By maintaining a fully analytic framework, we go beyond standard approximation schemes, typically limited to ultralight fields or weakly rotating and weakly charged regimes, and explicitly reveal how the four independent charges of the dyonic Kerr-Sen black hole shape the quasi-stationary spectrum. \\

\section{The Rotating Black Hole with Dilaton and Axion Charges}
The dyonic Kerr-Sen black hole represents a rotating and charged solution of the low-energy effective action of heterotic string theory. It extends the Kerr geometry by incorporating nontrivial dilaton and axion fields, which modify both the causal structure of the spacetime and the propagation of test fields.

In Boyer-Lindquist coordinates $(t,r,\theta,\phi)$, the spacetime line element is given by \cite{Wu_2021}
\begin{widetext}
\begin{multline}
ds^2 = - \left[ 1 - \frac{r_s (r-d) - r_D^2}{\rho^2} \right] c^2 dt^2- \frac{2 \left[r_s (r-d) - r_D^2\right]}{\rho^2} a \sin^2\theta d\phi c dt \\+ \frac{\rho^2}{\Delta} dr^2 + \rho^2 d\theta^2+ \left[ r(r-2d) - k^2 + a^2
+ \frac{a^2 \sin^2\theta \left[r_s (r-d) - r_D^2\right]}{\rho^2} \right]
\sin^2\theta d\phi^2 ,
\end{multline}  
\end{widetext}
where the metric functions are defined as
\begin{align}
\rho^2 &= r(r-2d) - k^2 + a^2 \cos^2\theta , \\
\Delta &= r(r-2d) - r_s (r-d) - k^2 + a^2 + r_D^2 \nonumber\\ 
&\equiv (r-r_+)(r-r_-).
\end{align}
Here, $r_+$ and $r_-$ denote the outer and inner horizons respectively, given by
\begin{equation}
 r_{\pm }=\frac{r_s}{2}+d\pm \sqrt{{\left(\frac{r_s}{2}\right)}^2+d^2+k^2-\left(a^2+r^2_D\right)}.
\end{equation}

The parameters entering the solution are defined as
\begin{align}
r_s = \frac{2GM}{c^2}, \qquad
r_D^2 = Q^2 + P^2, \\
k = \frac{2PQ}{r_s}, \qquad
d = \frac{P^2 - Q^2}{r_s},
\end{align}
where $M$ is the black hole mass, and $Q$ and $P$ represent the electric and magnetic charges, respectively.

There is a {\it toroidal} singularity in the DKS spacetime in these coordinates at $\rho^2 =0$,
\begin{equation}
    r_{p,m}=d \pm \sqrt{d^{2}+k^{2}-a^{2}\cos^{2}\theta},
\end{equation}
where both signs of $r_{p,m}$ actually form a single toroidal surface in the toroidal coordinates $(x,y,z)$ given by
\begin{align}
    x&=\sqrt{(r-d)^{2}+a^{2}}\sin\theta\cos\phi, \notag\\
    y&=\sqrt{(r-d)^{2}+a^{2}}\sin\theta\sin\phi, \notag\\
    z&=(r-d)\cos\theta, 
\end{align}
where $R\equiv\sqrt{x^{2}+y^{2}+z^{2}}=d^{2}+k^{2}-a^{2}\cos 2\theta$. 

The toroidal singularity could truncate into a spherical surface when $a=0$~(non-rotating). In the rotating case, we can find a path that evades the toroidal singularity for $R<d^{2}+k^{2}-a^{2}$ and $R>\sqrt{d^{2}+k^{2}+a^{2}}$. In the non-rotating case, the singularity becomes a spherical surface which causally separates the spacetime region between $R<\sqrt{d^{2}+k^{2}}$ and $R>\sqrt{d^{2}+k^{2}}$. 

\subsection{Horizon-Regular Coordinate}
To construct coordinates that remain regular across the event horizon, we introduce the tortoise coordinate $r_*$, defined by
\begin{equation}
\frac{dr_*}{dr}
= \frac{r(r-2d) - k^2 + a^2}{\Delta}.
\end{equation}
This coordinate is essential for describing the near-horizon region and will be used in the analysis of wave propagation.

Using $r_*$, we define the ingoing Eddington-Finkelstein coordinate
\begin{equation}
v = t + r_*,
\end{equation}
which is regular on the future horizon. 

Differentiating this relation gives
\begin{equation}
dv = dt + \frac{r(r-2d) - k^2 + a^2}{\Delta} dr,
\end{equation}
and hence
\begin{equation}
dt = dv - \frac{r(r-2d) - k^2 + a^2}{\Delta} dr .
\end{equation}

We introduce a redefinition of the azimuthal coordinate
\begin{equation}
d\tilde{\phi}=d\phi+\frac{a}{\Delta}dr,
\end{equation}
which implies
\begin{equation}
d\phi=d\tilde{\phi}-\frac{a}{\Delta}dr.
\end{equation}

Substituting these transformations into the Boyer-Lindquist line element, the metric can be expressed in ingoing Eddington-Finkelstein coordinates $(v,r,\theta,\tilde{\phi})$ as
\begin{widetext}
\begin{multline}
ds^2 = -\left(1-\frac{r_s (r-d)-r_D^2}{\rho^2}\right) dv^2 + 2\, dv\, dr - 2 a \sin^2\theta\, dr\, d\tilde{\phi} 
- 2 \frac{a \sin^2\theta\left[r_s (r-d)-r_D^2\right]}{\rho^2}\, dv\, d\tilde{\phi} + \\\rho^2 d\theta^2 
+ \sin^2\theta \left( r(r-2d)-k^2+a^2 + \frac{a^2 \sin^2\theta\left[r_s (r-d)-r_D^2\right]}{\rho^2} \right) d\tilde{\phi}^2 .
\end{multline}
 
\end{widetext}

\section{Relativistic Massive Wave Dynamics}
The evolution of a scalar field $\psi$ with mass $m$ in a curved spacetime is described by the covariant Klein-Gordon equation
\begin{equation}
\left[ - \frac{1}{\sqrt{-g}} \partial_\mu
\left( \sqrt{-g} g^{\mu\nu} \partial_\nu \right)
+ m^2 \right]\psi = 0, \label{KGeq}
\end{equation}

\noindent which constitutes the master equation governing scalar perturbations in the Kerr-Sen background. This equation tells us that the propagation of the scalar field is controlled by both its intrinsic mass and the spacetime geometry, leading to a nontrivial coupling between the field parameters $(m,\omega,m_l)$ and the black hole parameters $(M,a,Q,P)$.

The Laplace-Beltrami operator appearing in Eq.~\eqref{KGeq} can be evaluated component by component as
\begin{widetext}
    \begin{gather}
\frac{1}{\sqrt{-g}} \partial_v \left( \sqrt{-g} g^{vv}\partial_v\psi \right) = \frac{a^2\sin^2\theta}{\rho^2}\,\partial_v^2\psi , \\ 
\frac{1}{\sqrt{-g}}\partial_v \left(\sqrt{-g}g^{vr}\partial_r\psi\right) +\frac{1}{\sqrt{-g}}\partial_r \left(\sqrt{-g}g^{rv}\partial_v\psi\right) = \frac{2A}{\rho^2}\,\partial_v\partial_r\psi + \frac{\partial_r A}{\rho^2}\,\partial_v\psi , \\
\frac{1}{\sqrt{-g}}\partial_r \left(\sqrt{-g}g^{rr}\partial_r\psi\right) = \frac{\Delta}{\rho^2}\,\partial_r^2\psi + \frac{\partial_r\Delta}{\rho^2}\,\partial_r\psi ,\\
\frac{1}{\sqrt{-g}}\partial_\theta \left(\sqrt{-g}g^{\theta\theta}\partial_\theta\psi\right) = \frac{1}{\rho^2} \left( \partial_\theta^2\psi + \cot\theta\,\partial_\theta\psi \right),\\
\frac{1}{\sqrt{-g}} \partial_v \left( \sqrt{-g}g^{v\phi}\partial_{\tilde\phi}\psi \right) + \frac{1}{\sqrt{-g}} \partial_{\tilde\phi} \left( \sqrt{-g}g^{\phi v}\partial_v\psi \right) = \frac{2a}{\rho^2}\,\partial_v\partial_{\tilde\phi}\psi , 
\end{gather}
\begin{gather}
\frac{1}{\sqrt{-g}} \partial_r \left( \sqrt{-g}g^{r\phi}\partial_{\tilde\phi}\psi \right) + \frac{1}{\sqrt{-g}} \partial_{\tilde\phi} \left( \sqrt{-g}g^{\phi r}\partial_r\psi \right) = \frac{2a}{\rho^2}\,\partial_r\partial_{\tilde\phi}\psi , \\ 
\frac{1}{\sqrt{-g}}\partial_{\tilde\phi} \left(\sqrt{-g}g^{\phi\phi}\partial_{\tilde\phi}\psi\right) = \frac{1}{\rho^2\sin^2\theta}\,\partial_{\tilde\phi}^2\psi .
\end{gather}
\end{widetext}

By combining these contributions, we obtain the explicit form of the Laplace-Beltrami operator
\begin{multline}
\frac{1}{\sqrt{-g}} \partial_\mu \left( \sqrt{-g} g^{\mu\nu} \partial_\nu \right)\psi =\\ \frac{1}{\rho^2}\Bigg[ a^2 \sin^2\theta\,\partial_v^2 + 2A(r)\,\partial_v\partial_r + (\partial_r A)\,\partial_v\\
+ \partial_r \bigl(\Delta\,\partial_r\bigr) + 2a\,(\partial_v+\partial_r)\partial_{\tilde\phi} + \nabla_{S^2}^2 \Bigg]\psi,
\end{multline}
where the angular Laplacian is defined as
\begin{equation} 
\nabla_{S^2}^2 \equiv \frac{1}{\sin\theta}\partial_\theta \left(\sin\theta\partial_\theta \right) + \csc^2\theta\,\partial_{\tilde\phi}^2 . 
\end{equation}

We adopt the separable ansatz
\begin{equation}
\psi(v,r,\theta,\tilde \phi) = e^{-i\omega v} e^{i m_l \tilde\phi} R(r) S(\theta),
\end{equation}
which reflects the axial symmetry of the spacetime. In this form, the frequency $\omega$ and azimuthal number $m_l$ are associated with conserved quantities arising from the Killing vectors $\partial_v$ and $\partial_{\tilde\phi}$.

Substituting this ansatz into Eq.~\eqref{KGeq}, we find that the Klein-Gordon equation separates into radial and angular parts. After straightforward algebraic manipulation, we obtain
\begin{widetext}
    \begin{multline}
\underbrace{ \frac{1}{R}\partial_r \left(\Delta\,\partial_r R\right) -2 i\bigl(\omega A-a m_l\bigr)\frac{\partial_r R}{R} - (r(r-2d) - k^2)\, m^2 + 2a\omega m_l -2i\omega(r-d)- a^2\omega^2 }_{\text{radial part}} \\
+ \underbrace{ \frac{1}{S} \left( \frac{1}{\sin\theta}\partial_\theta \left(\sin\theta\partial_\theta S\right) -\frac{m_l^2}{\sin^2\theta} S \right) -a^2(m^2-\omega^2)\cos^2\theta }_{\text{angular part}} =0 .
\end{multline}
\end{widetext}

\subsection{Angular equation}
The angular dependence separates completely and reduces to a second-order ordinary differential equation. Writing the operator in self-adjoint form, we obtain \cite{Press,Berti1,Berti2,Cho:2009wf,Hod:2015cqa,Ponglertsakul:2020ufm}
\begin{multline}
\frac{1}{\sin\theta}\frac{d}{d\theta}\!\left(\sin\theta \frac{dT}{d\theta}\right) -\frac{m_l^2}{\sin^2\theta}\,T \\+ a^2\!\left(\omega^2-m^2\right)\cos^2\theta\,T = -\lambda\,T,
\end{multline}
where $\lambda$ is the angular separation constant.

We note that the parameter $\sigma=a^2(\omega^2-m^2)$ controls the deviation from spherical symmetry, and therefore quantifies the influence of rotation on the angular distribution of the scalar field.

For small spheroidicity, the eigenvalue $\lambda$ admits the perturbative expansion
\begin{equation}
\lambda^{m_l}_l = l(l+1) + 2\sigma \left(\frac{m_l^2 + l(l+1) - 1}{(2l-1)(2l+3)}\right) + \mathcal{O}(\sigma^2),
\end{equation}
with\\
\begin{equation}
\sigma = a^2\left(\omega^2-m^2\right).
\end{equation}

In the non-rotating limit $a=0$, the equation reduces to that of spherical harmonics, with eigenvalues
\begin{equation}
\lambda = l(l+1),
\end{equation}
and eigenfunctions given by the associated Legendre polynomials $P_l^{m_l}(\cos\theta)$.

For $a \neq 0$, the $\cos^2\theta$ term deforms the angular equation, and the solutions are given by spheroidal harmonics, reflecting the imprint of rotation on the angular structure of the scalar field. The angular function can be written as
\begin{align}
T(\theta)&= S^{m_l}_l\!\left(\sigma,\cos\theta\right) \nonumber \\
&= \sum_{r=-\infty}^{\infty} d^{l m_l}_r(\sigma)\, P^{m_l}_{l+r}(\cos\theta).
\end{align}

\subsection{Radial equation}
After separating the angular dependence, the scalar field dynamics reduce to the following radial equation
\begin{multline}
 \partial_r \left(\Delta\,\partial_r R\right)-2 i\bigl(\omega A-a m_l\bigr)\partial_r R\\+\left[ - (r(r-2d) - k^2)\, m^2 +2a\omega m_l\right. \\ \left. -2i\omega(r-d)-a^2\omega^2-\lambda\right]R=0. \label{radialeq}
\end{multline}

Let us introduce a dimensionless radial coordinate defined by\\
\begin{gather}
\delta_r = r_+ - r_- ,\\
    \delta_r z = r - r_- , \qquad dr = \delta_r dz ,\\
r - r_+ = \delta_r (z - 1) ,
\end{gather}
which maps the interval between the horizons to $0 \le z \le 1$. This transformation makes the singular structure of the differential equation manifest at $z=0$ and $z=1$, corresponding to the inner and outer horizons, respectively.

In terms of the new coordinate $z$, the radial equation becomes
\begin{widetext}
    \begin{multline}
z(z-1)\,\partial_z^2 R(z) -\left[(1-2z)  + \frac{2 i }{\delta_r}\Big\{\big(a^2 - k^2 + (r_- + z\delta_r)(r_- + z\delta_r - 2d)\big)\omega- a m_l \Big\} \right]\partial_z R(z) \\ -\left[ m^2 \Big\{- k^2 + (r_- + z\delta_r)(r_- + z\delta_r - 2d)\Big\}  - 2 a m_l \omega + 2 i (r_- + z\delta_r - d)\omega +a^2\omega^2 + \lambda \right]R(z)=0 . \label{fulleq}
\end{multline}
\end{widetext}

By comparing Eq.~\eqref{fulleq} with the general second-order differential equation \eqref{general ODE}, we identify the coefficient functions
\begin{multline}
p(z) = -\frac{1}{z (z-1)} \left[(1-2z) +\frac{2 i }{\delta_r} \times\right. \\ \left. \Big\{\big(a^2 - k^2 + (r_- + z\delta_r)(r_- + z\delta_r - 2d)\big)\omega- a m_l \Big\}\right],
\end{multline}
\begin{multline}
    q(z) =- \frac{1}{z (z-1)} \left[ \lambda - 2 a m_l \omega + 2 i (r_- + z\delta_r - d)\omega  \right. \\ \left.  +a^2\omega^2+ m^2 \Big\{- k^2 + (r_- + z\delta_r)(r_- + z\delta_r - 2d)\Big\}\right].
\end{multline}

Following Appendix \ref{AppendixA}, we remove the first-derivative term through the transformation
\begin{multline}
R(z) = Z(z) e^{-\frac{1}{2} \int p(z) dz}= Z(z)e^{i\delta_r\omega z}\times\\ (1-z)^{-\frac12-\frac{i}{\delta_r}\left[ a m_l-\omega\left(a^{2}-k^{2}+r_+(r_+ - 2d)\right) \right]} \times \\z^{-\frac12+\frac{i}{\delta_r}\left[ a m_l-\omega\left(a^{2}-k^{2}+r_-(r_- - 2d)\right) \right]} .
\label{RtoZ}
\end{multline}

This transformation brings the radial equation into its normal form
\begin{equation}
\partial_z^2 Z(z) + K(z) Z(z) = 0 ,
\end{equation}
where the effective potential $K(z)$ is given by
\begin{equation}
K(z) = - \delta_r^2 ( m^2 - \omega^2 ) + \frac{K_1}{z} + \frac{K_2}{z^2} + \frac{K_3}{z-1} + \frac{K_4}{(z-1)^2}.
\end{equation}
\\
The structure of $K(z)$ shows that the singular points at $z=0$ and $z=1$ are regular, confirming that the equation belongs to the confluent Heun class. The coefficients $K_i$ encode the full dependence on the black hole charges and rotation, as well as on the scalar field parameters.

By comparing this form with the standard confluent Heun equation (see Appendix \ref{AppendixB}), we identify the corresponding Heun parameters
\begin{gather}
\alpha_\pm = \pm 2 \delta_r \sqrt{ m^2 - \omega^2 }, \label{alpha} \\ 
\beta_\pm = \pm \frac{2 i}{\delta_r} \left[ a m_l -\omega\left(a^{2}-k^{2}+r_-(r_- - 2d)\right) \right], \label{beta} \\
\gamma_\pm = \pm \frac{2 i}{\delta_r} \left[ a m_l -\omega\left[a^{2}-k^{2}+r_+(r_+-2d)\right] \right], \label{gamma}\\
\delta = {-\delta_r r_s \left( m^2 - 2\omega^2\right) }, \\ 
\eta = \frac{1}{2} - K_1 .
\end{gather}

With these identifications, the general solution of the normal form equation can be expressed in terms of the confluent Heun function as
\begin{multline}
Z(z) = Z_0 e^{ \frac{1}{2} \alpha_\pm z } (z-1)^{ \frac{1}{2} ( \gamma_\pm + 1 ) } z^{ \frac{1}{2} ( \beta_\pm + 1 ) } \times \\\operatorname{HeunC} \left( \alpha_\pm , \beta_\pm , \gamma_\pm , \delta , \eta , z \right) .
\end{multline}

Combining this expression with Eq.~\eqref{RtoZ}, we obtain the exact radial solution in closed form
\begin{multline}
R(z)= R_0 e^{\frac12 \alpha_\pm z + i \delta_r \omega z} z^{\frac12(\beta_\pm+\beta_+)}(z-1)^{\frac12(\gamma_\pm-\gamma_+)}\times \\ \operatorname{HeunC} \left(\alpha_\pm,\beta_\pm,\gamma_\pm,\delta,\eta,z\right),
\end{multline}
where the dimensionless radial coordinate is given by
\begin{equation}
z = \frac{ r - r_- }{ \delta_r } .
\end{equation}

\onecolumngrid

\begin{figure}[h]
    \centering
    \includegraphics[scale=0.65]{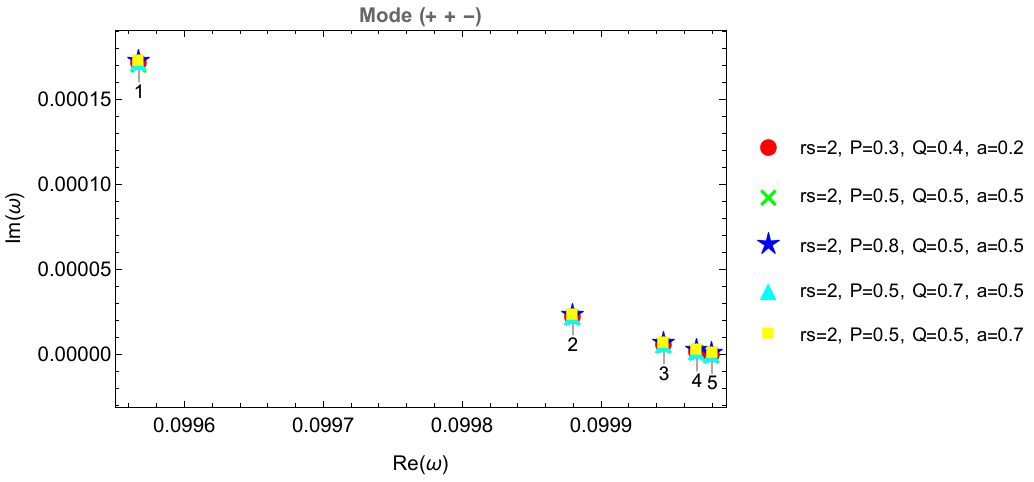}\\
    \includegraphics[scale=0.65]{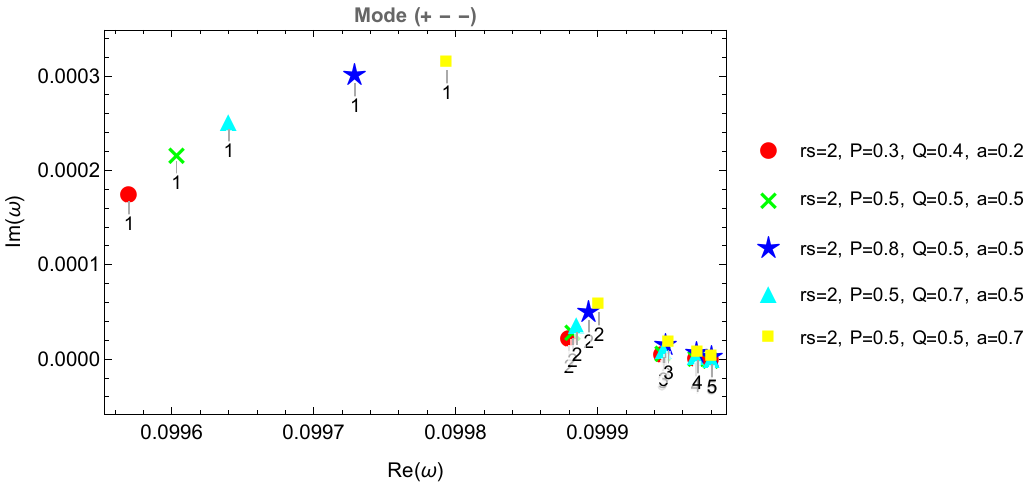}\\
   \caption{Quasi-stationary spectrum in the complex frequency plane for the dyonic Kerr-Sen black hole with $m=0.1$ and $m_\ell=0$.}
   \label{QBSa}
\end{figure}

\begin{figure}[h]
    \centering
     \includegraphics[scale=0.65]{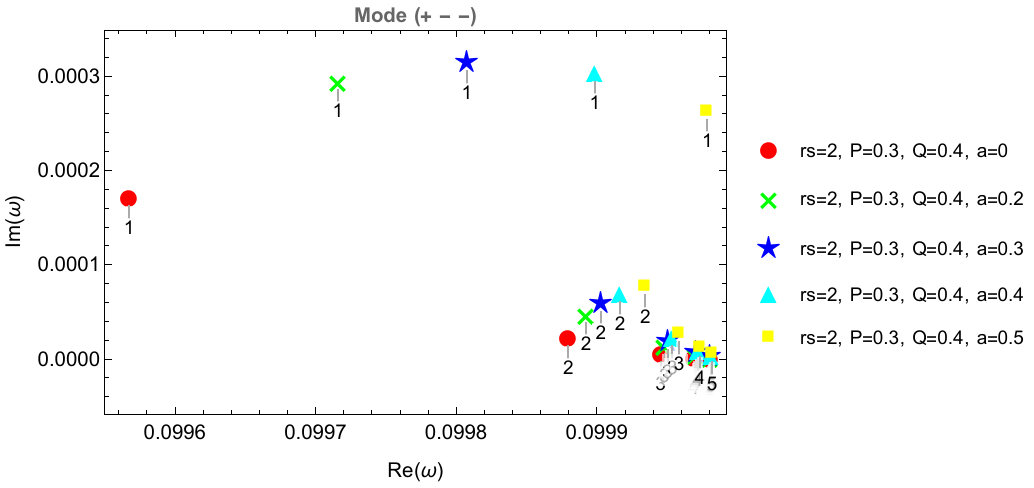}
   \caption{Quasi-stationary spectrum for counter-rotating modes ($m_\ell=-1$) with $P=0.3$, $Q=0.4$, and $m=0.1$.}
   \label{QBS-1}
\end{figure}

\begin{figure}[h]
    \centering
    \includegraphics[scale=0.65]{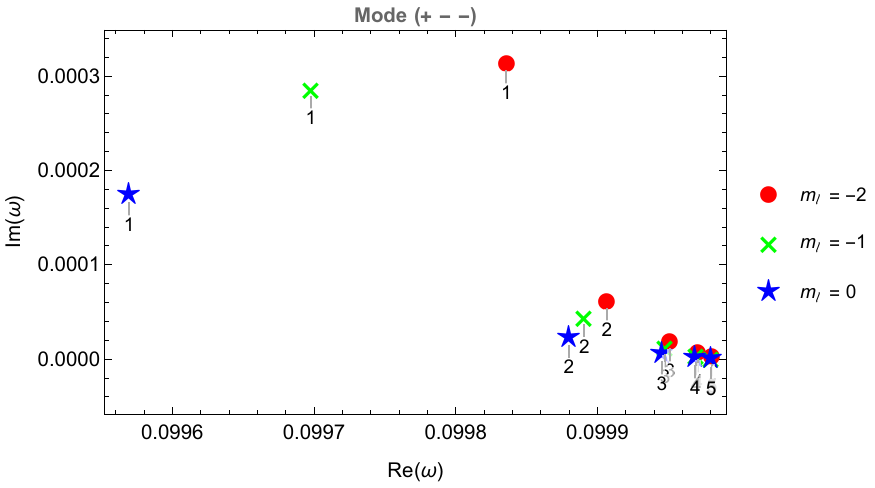}
   \caption{Dependence of the quasi-stationary spectrum on the magnetic quantum number $m_\ell$ for $a=0.2$, $P=0.3$, $Q=0.4$, and $m=0.1$.}
   \label{QBSml}
\end{figure}

\begin{figure}[h]
    \centering
    \includegraphics[scale=0.65]{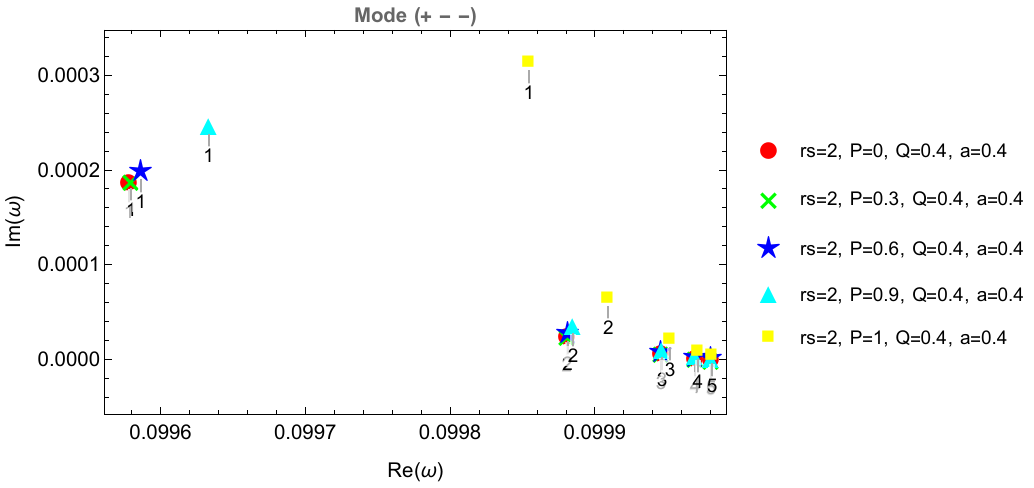}
   \caption{Quasi-stationary spectrum of $m=0.1$ scalar field for $m_\ell=0$, $a=0.4$, and $Q=0.4$.}
   \label{QBSP}
\end{figure}

\begin{figure}[h]
    \centering
    \includegraphics[scale=0.65]{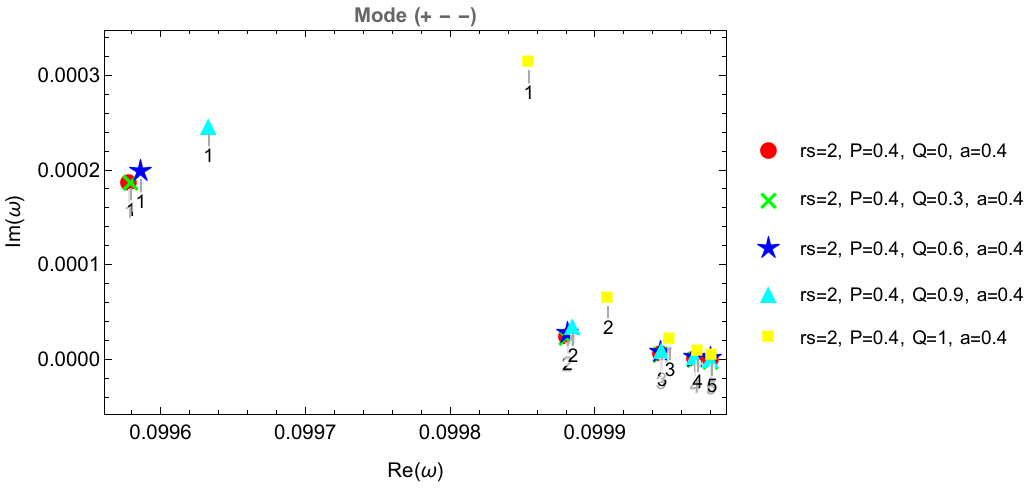}
   \caption{Quasi-stationary spectrum of $m=0.1$ scalar field for $m_\ell=0$, $a=0.4$, and $P=0.4$.}
   \label{QBSQ}
\end{figure}

\twocolumngrid

\subsubsection{Near-inner-horizon behavior}
In the vicinity of the inner horizon $r=r_-$, corresponding to $z\to 0$, the radial solution behaves as
\begin{equation}
R(z) \sim z^{\frac12(\beta_\pm+\beta_+)} .
\end{equation}

For the $\beta_-$ branch, we obtain
\begin{equation}
R_{\beta_-}(z) \sim z^{\frac12(\beta_-+\beta_+)} = z^{0},
\end{equation}
which is finite and non-oscillatory at the inner horizon. This behavior indicates that the mode remains regular at $r=r_-$ and corresponds to a purely ingoing wave in the near-horizon region.

For the $\beta_+$ branch, the solution behaves as
\begin{equation}
R_{\beta_+}(z) \sim z^{\beta_+},
\end{equation}
which is oscillatory due to the imaginary nature of $\beta_+$. This mode describes wave propagation near the inner horizon, with its ingoing or outgoing character determined by the associated Klein-Gordon flux.

\subsubsection{Near-outer-horizon behavior}
In the vicinity of the outer (event) horizon $r=r_+$, corresponding to $z\to1$, the radial solution behaves as
\begin{equation}
R(z) \sim (z-1)^{\frac12(\gamma_\pm -\gamma_+)} .
\end{equation}

For the $\gamma_-$ branch, we find
\begin{equation}
R_{\gamma_-}(z) \sim (z-1)^{-\gamma_+},
\end{equation}
which represents an oscillatory mode corresponding to outgoing radiation at the event horizon.

For the $\gamma_+$ branch, the solution reduces to
\begin{equation}
R_{\gamma_+}(z) \sim (z-1)^{0},
\end{equation}
which is finite and non-oscillatory. This branch corresponds to a purely ingoing mode at the event horizon.

\subsubsection{Inside the new universe}
At spatial infinity ($r \to -\infty$), the radial solution exhibits the asymptotic behavior
\begin{equation}
R(r \to -\infty) \sim e^{\frac12 \alpha_\pm z}.
\end{equation}

In the maximally extended spacetime, the limit $r \to -\infty$ is interpreted as a second asymptotically flat exterior region, corresponding to a ``new universe'' connected through the black hole interior. In this second universe, we require exponentially decaying solutions as $r \to -\infty$, corresponding to the $\alpha_{+}$ branch of the spectrum. Moreover, since $\operatorname{Re}(\alpha_+)>0$, these modes also satisfy the outgoing-wave condition in the limit $r \to -\infty$. Such modes are physically admissible, since no incoming wave should emerge from the new universe into the black hole interior.

\subsection{Energy quantization}
The discrete quasi-stationary spectrum arises from the boundary conditions required for a trapped scalar configuration around the black hole. In particular, a quasi-stationary mode must not blow up at spatial infinity. The general radial solution, however, does not automatically satisfy this requirement. It is typically expressed as an infinite series, which in general includes terms that grow at large radius and therefore spoil localization. To obtain physically acceptable solutions, these divergent contributions must be eliminated. This is achieved by requiring the series to terminate, reducing it to a polynomial of finite degree. This termination condition directly yields the quantization condition for the spectrum. As a result, the allowed frequencies are no longer continuous, but instead form a discrete spectrum.

For the confluent Heun function, the series termination condition is given by \cite{Heun,Fiziev_2009},
\begin{align}
\frac{\delta}{\alpha} + \frac{\beta + \gamma}{2} = -n , \label{quanfor}
\end{align}
where $n = n_r + 1 = 1,2,\ldots$ labels the radial quantum number.

As a consequence, Eq.~\eqref{quanfor} acts as a quantization condition on the scalar-field frequency. Only specific discrete values satisfy this relation, leading to a quasi-stationary spectrum. In this sense, the radial quantum number $n_r$ plays a role analogous to that in familiar bound state problems, corresponding to the number of nodes in the radial profile and organizing the spectrum into a ladder of states.

Now, let us consider the quantization condition \eqref{quanfor} for different sign choices of the parameters $\alpha$, $\beta$, and $\gamma$. Each branch leads to a distinct algebraic relation for the frequency $\omega$, indicating different couplings between the scalar field and the black hole parameters.

\begin{enumerate}
\item For the branch $(\alpha_+,\beta_+,\gamma_+)$, we obtain
\begin{equation}
-\frac{ (m^2-2\omega^2) r_s } {2 \sqrt{m^2-\omega^2}}+\frac{i}{\delta_r} \left[2 a m_l-(r_s^2-2 r_D^2)\omega\right] = -n . \label{1}
\end{equation}
This relation depends explicitly on the rotation parameter $a$ and the total charge scale $r_D$, showing that both rotation and electromagnetic charges directly influence the energy and stability of the scalar field.

\item For the branch $(\alpha_+,\beta_+,\gamma_-)$, we find
\begin{equation}
-\frac{ (m^2-2\omega^2) r_s} {2 \sqrt{m^2-\omega^2}} + i r_s \omega = -n . \label{2}
\end{equation}
This branch is independent of $a$, $Q$, and $P$, indicating that the corresponding modes are governed purely by the scalar mass and the overall gravitational scale set by the black hole.

\item For the branch $(\alpha_+,\beta_-,\gamma_+)$, we obtain
\begin{equation}
-\frac{ (m^2-2\omega^2) r_s } {2 \sqrt{m^2-\omega^2}} - i r_s \omega = -n . \label{3}
\end{equation}
This branch is also independent of rotation and charges, but differs in the sign of the imaginary term, which controls whether the mode decays or grows.

\item Finally, for the branch $(\alpha_+,\beta_-,\gamma_-)$, we obtain
\begin{equation}
-\frac{ (m^2-2\omega^2) r_s} {2 \sqrt{m^2-\omega^2}}-\frac{i}{\delta_r} \left[ 2 a m_l-(r_s^2-2 r_D^2)\omega\right] = -n , \label{4}
\end{equation}
\end{enumerate}
where we have used the identity
\begin{equation}
2 d - 2 r_- - \delta_r = - r_s. \\
\end{equation}

Since we can cast $\displaystyle{\delta_{r}=2\sqrt{\left(\frac{r_{s}}{2}-\frac{r_{D}^{2}}{r_{s}}\right)^{2}-a^{2}}}$, all of the quasi-stationary states above depend only on $r_{s},a,m$ and the charge length scale $r_{D}$. 

Each quantization condition yields a quartic equation for the frequency $\omega$ and numerical analysis indicates that the generic roots are either complex or purely imaginary. The complex roots correspond to propagating quasi-stationary modes with oscillatory behavior accompanied by amplification or damping, whereas the purely imaginary roots describe non-propagating configurations that evolve exponentially in time without oscillatory propagation.

Table~\ref{tab:quasi_freq} displays the quasi-stationary frequencies for the different sign configurations in the ground state, where $m = 0.1, r_s = 1, P = 0.1, Q = 0.1, a = 0.1$. For the $(+,+,+)$ and $(+,-,-)$ branches, four independent roots are obtained. In both cases, two of the frequencies are purely imaginary, while the remaining pair forms a symmetric complex structure with opposite real parts and identical imaginary contributions. Moreover, the spectrum associated with the $(+,-,-)$ branch is the complex conjugate of the $(+,+,+)$ spectrum. The $(+,+,-)$ and $(+,-,+)$ branches, on the other hand, admit only three distinct roots. The spectra of the $(+,+,-)$ and $(+,-,+)$ configurations are also related through complex conjugation.  

Table~\ref{tab:alpha_values} displays the corresponding values of the parameter $\alpha$ associated with the quasi-stationary frequencies shown in Table~\ref{tab:quasi_freq}. As in the frequency spectrum, the $(+,+,+)$ and $(+,-,-)$ branches admit four distinct roots, consisting of two purely real solutions together with a pair of complex conjugate roots. In contrast, the $(+,+,-)$ and $(+,-,+)$ branches possess only three roots, namely one purely real solution and a pair of complex conjugate roots. Furthermore, the spectra of the $(+,-,-)$ and $(+,+,+)$ branches are related through complex conjugation, while the same relation holds between the $(+,-,+)$ and $(+,+,-)$ configurations.

Table~\ref{tab:quasi_freqml+1} and \ref{tab:quasi_freqml-1} present the non-spherical quasi-stationary frequencies for the various sign configurations in the ground state, with parameters $m = 0.1$, $r_s = 1$, $P = 0.1$, $Q = 0.1$, and $a = 0.1$. As in the spherical case, the spectrum associated with the $(+,-,-)$ branch is the complex conjugate of the $(+,+,+)$ spectrum. In contrast, the $(+,+,-)$ and $(+,-,+)$ branches admit only three distinct roots, and their spectra are likewise related by complex conjugation.

The asymptotic behavior of the radial wave function is given by
\begin{equation}
   \psi \sim e^{-i\omega t}e^{\frac{\alpha_+}{2\delta_r}r},
\qquad \delta_r>0, \label{asymp}
\end{equation}
where both $\omega$ and $\alpha_+$ are, in general, complex quantities. Writing $ \omega=\omega_R+i\omega_I$ and $\alpha=\alpha_R+i\alpha_I$, the wave function takes the form
\begin{equation}
\psi \sim e^{\omega_I t+\frac{\alpha_{+R}}{2\delta_r}r}
\,e^{-i\left(\omega_R t-\frac{\alpha_{+I}}{2\delta_r}r\right)}  
\end{equation}

The character of the quasi-stationary modes can be determined from the conserved Klein-Gordon current,
\begin{equation}
j^\mu=\frac{1}{2mi}\left(\psi^\ast \nabla^\mu\psi -\psi \nabla^\mu\psi^\ast \right).
\end{equation}

Using \eqref{asymp} the radial component of the current behaves asymptotically as
\begin{equation}
j^r=\frac{ \alpha_{+I}}{2m\delta_r}\, e^{2\omega_I t+\frac{\alpha_{+R}}{\delta_r}r}.
\end{equation}

The sign of $j^r$ determines the direction of the radial flux. In particular, $j^r<0$ corresponds to an ingoing mode, while $j^r>0$ describes an outgoing mode. Therefore, imposing the outgoing boundary condition at
\begin{equation}
r\rightarrow -\infty
\end{equation}
requires $\mathrm{Im}(\alpha)<0$.

In addition, regularity of the radial wave function demands that the exponential factor remains finite as $r\to -\infty$, which implies $\mathrm{Re}(\alpha)>0$. Consequently, the physically admissible quasi-stationary branch is characterized by
\begin{equation}
\mathrm{Re}(\alpha)>0,\qquad \mathrm{Im}(\alpha)<0,
\end{equation}
and this corresponds to the root of $\omega$ with the following characteristics: modes with positive real frequency grow exponentially in time, whereas those with negative real frequency decay.  

In Fig.~\ref{QBSa}, we depict the physical quasi-stationary spectrum satisfying $\omega_R>0$ for fixed scalar-field parameters $m=0.1$ and $m_\ell=0$. An interesting feature emerges from the behavior of the $(\alpha_{+},\beta_{+},\gamma_{-})$ branch, which exhibits no dependence on either the black hole rotation parameter or the electromagnetic charges. The corresponding mode is largely insensitive to the detailed structure of the rotating charged spacetime and is governed primarily by the scalar mass and the overall gravitational scale.

\onecolumngrid

\begin{table}[ht]
\centering
\renewcommand{\arraystretch}{1.2}
\begin{tabular}{|c|c|c|c|c|}
\hline
\multicolumn{5}{|c|}{\textbf{Quasi-Stationary Spectrum for Different Sign Configurations}} \\ \hline
Configuration & $\omega_{1}$ & $\omega_{2}$ & $\omega_{3}$ & $\omega_{4}$ \\ \hline
$(+,+,+)$ & $-0.0998799 - 0.000024561\, i$ & $-0.494404\, i$ & $-44.5745\, i$ & $0.0998799 - 0.000024561\, i$ \\ \hline
$(+,+,-)$ & $+0.499952\, i$ & $-0.0998797 + 0.0000240429\, i$ & $0.0998797 + 0.0000240429\, i$ & --- \\ \hline
$(+,-,+)$ & $-0.499952\, i$ & $-0.0998797 - 0.0000240429\, i$ & $0.0998797 - 0.0000240429\, i$ & --- \\ \hline
$(+,-,-)$ & $-0.0998799 + 0.000024561\, i$ & $+0.494404\, i$ & $+44.5745\, i$ & $0.0998799 + 0.000024561\, i$ \\ \hline
\end{tabular}
\caption{Quasi-stationary frequencies of states $n=1,m_{l}=0$ for different sign configurations for $m = 0.1, r_s = 1, P = 0.1, Q = 0.1$ and $a = 0.1$.}
\label{tab:quasi_freq}
\end{table}

\begin{table}[ht]
\centering
\renewcommand{\arraystretch}{1.2}
\begin{tabular}{|c|c|c|c|c|}
\hline
\multicolumn{5}{|c|}{\textbf{Values of $\alpha$ Associated with the Quasi-Stationary Modes}} \\ \hline
Configuration & $\alpha_{1}$ & $\alpha_{2}$ & $\alpha_{3}$ & $\alpha_{4}$ \\ \hline
$(+,+,+)$ & $0.00924898 - 0.000935322 \, i$ & $0.947229$ & $83.7053 $ & $0.00924898 + 0.000935322\, i$ \\ \hline
$(+,+,-)$ & $0.957442 $ & $0.00925302 + 0.000915192 \, i$ & $0.00925302 - 0.000915192 \, i$ & --- \\ \hline
$(+,-,+)$ & $0.957442 $ & $0.00925302 - 0.000915192 \, i$ & $0.00925302 + 0.000915192 \, i$ & --- \\ \hline
$(+,-,-)$ & $0.00924898 + 0.000935322\, i$ & $0.947229 $ & $83.7053$ & $0.00924898 - 0.000935322 \, i$ \\ \hline
\end{tabular}
\caption{Values of the parameter $\alpha$ for each configuration presented in Table~\ref{tab:quasi_freq}.}
\label{tab:alpha_values}
\end{table}

\twocolumngrid 

By contrast, the $(\alpha_{+},\beta_{-},\gamma_{-})$ branch depends explicitly on both the rotation parameter $a$ and the effective charge scale $r_D$, signaling a nontrivial coupling between the scalar field and the rotating charged spacetime. The resulting spectrum therefore encodes direct information about the background geometry and its electromagnetic structure.

Fig.~\ref{QBS-1} displays frequency profiles of non-spherical configurations of the counter-rotating $(\alpha_{+},\beta_{-},\gamma_{-})$ branch. Furthermore, Fig.~\ref{QBSml} illustrates the behavior of the spectrum for several values of the azimuthal number $m_\ell$, while keeping the scalar mass and black hole parameters fixed. As in the spherical modes, the particle modes generically possess positive imaginary frequencies and therefore exhibit an unstable time evolution.

Figures~\ref{QBSP} and \ref{QBSQ} further reveal that increasing either the electric charge $Q$ or the magnetic charge $P$ produces a symmetric shift of the spectrum in the complex-frequency plane. This is a consequence of the dependence on only $r_{s},a,m$ and the charge length scale $r_{D}=\sqrt{P^{2}+Q^{2}}$ of the quasi-stationary states. In particular, the real part $\mathrm{Re}(\omega)$ increases with the increasing charges, indicating a higher oscillation frequency for the quasi-stationary modes. Simultaneously, the imaginary part, $\omega_I$, also grows, implying that the associated instability develops more rapidly as the electromagnetic charges become stronger.

\subsection{Probing the Chronology Protection Conjecture}
In this subsection, we briefly comment on Hawking's chronology protection conjecture (CPC), which states that the laws of physics prevent the formation of time machines or spacetime regions containing closed timelike curves (CTCs). Originally formulated by Hawking in 1992 \cite{Hawking:1991nk}, the conjecture is based on the observation that quantum fields propagating near a would-be chronology horizon undergo repeated blueshifts, signaling divergent backreaction and the onset of dynamical instabilities.

In rotating black holes, the region inside the inner horizon may contain CTCs, where particles or fields circling the causality-violating region are likewise expected to experience successive blueshifts, potentially destabilizing the geometry. In Hawking's semiclassical analysis of a traversable wormhole \cite{Hawking:1991nk}, a radiation beam entering one mouth was shown to realign through vacuum fluctuations before emerging from the other mouth, leading to an accumulation of energy sufficient to collapse the wormhole. Subsequent studies \cite{15,16} further demonstrated that bosonic perturbations of a wormhole throat can induce horizon bifurcation, resulting either in an inflationary universe or gravitational collapse into a black hole, depending on the sign of the injected energy. 

In our previous work \cite{Bunyaratavej:2024qgk}, we investigated the chronology protection conjecture through the spectrum of scalar quasi-stationary resonances. The analysis was performed in Boyer-Lindquist coordinates in the {\it inner region of the inner horizon}, where we selected the modes corresponding to outgoing solutions in the asymptotic region $r\to -\infty$, namely the sector associated with the closed-timelike-curve region. We found that all four outgoing branches exhibit the following behavior: modes with positive real frequency grow exponentially in time, whereas those with negative real frequency decay. This result suggests that whenever a massive scalar field penetrates the region containing closed timelike curves, physically relevant positive-frequency modes trigger instabilities capable of destabilizing the underlying spacetime geometry. Furthermore, the purely imaginary modes possess no oscillatory component and therefore do not propagate through the spacetime. Since these modes correspond only to exponentially damped or amplified configurations, they cannot support traveling excitations along closed timelike curves. In this sense, the purely imaginary sector does not lead to time-travel propagation and thus remains consistent with Hawking's Chronology Protection Conjecture (CPC).

\onecolumngrid

\begin{table}[ht]
\centering
\renewcommand{\arraystretch}{1.2}
\begin{tabular}{|c|c|c|c|c|}
\hline
\multicolumn{5}{|c|}{\textbf{Co-rotating Quasi-Stationary Spectrum }} \\ \hline
Configuration & $\omega_{1}$ & $\omega_{2}$ & $\omega_{3}$ & $\omega_{4}$ \\ \hline
$(+,+,+)$ & $-0.09997 - 9.33838\times 10^{-6}\, i$ & $0.09997 + 3.425\times 10^{-6} \, i$ & $0.105324 - 0.988901 \, i$ & $9.49468 - 89.1489 \, i$ \\ \hline
$(+,+,-)$ & $-0.09997 + 3.0941\times 10^{-6}\, i$ & $+0.999994 i$ & $0.09997 + 3.0941\times 10^{-6}\, i$ & --- \\ \hline
$(+,-,+)$ & $-0.09997 - 3.0941\times 10^{-6}\, i$ & $-0.999994 i$ & $0.09997 - 3.0941\times 10^{-6}\, i$ & --- \\ \hline
$(+,-,-)$ & $-0.09997 + 9.33838\times 10^{-6}\, i$ & $0.09997 - 3.425\times 10^{-6} \, i$ & $0.105324 + 0.988901 \, i$ & $9.49468 + 89.1489 \, i$ \\ \hline
\end{tabular}
\caption{Quasi-stationary frequencies of states $n=2,m_{l}=+1$ for different sign configurations for $m = 0.1, r_s = 1, P = 0.1, Q = 0.1$ and $a = 0.1$.}
\label{tab:quasi_freqml+1}
\end{table}

\begin{table}[ht]
\centering
\renewcommand{\arraystretch}{1.2}
\begin{tabular}{|c|c|c|c|c|}
\hline
\multicolumn{5}{|c|}{\textbf{Counter-rotating Quasi-Stationary Spectrum }} \\ \hline
Configuration & $\omega_{1}$ & $\omega_{2}$ & $\omega_{3}$ & $\omega_{4}$ \\ \hline
$(+,+,+)$ & $-9.49468 - 89.1489 \, i$ & $-0.105324 - 0.988901\, i$ & $-0.09997 + 3.425\times 10^{-6}\, i$ & $0.09997 - 9.33838\times 10^{-6} \, i$ \\ \hline
$(+,+,-)$ & $-0.09997 + 3.0941\times 10^{-6}\, i$ & $+0.999994 i$ & $0.09997 + 3.0941\times 10^{-6}\, i$ & --- \\ \hline
$(+,-,+)$ & $-0.09997 - 3.0941\times 10^{-6}\, i$ & $-0.999994 i$ & $0.09997 - 3.0941\times 10^{-6}\, i$ & --- \\ \hline
$(+,-,-)$ & $-9.49468 + 89.1489 \, i$ & $-0.105324 + 0.988901\, i$ & $-0.09997 - 3.425\times 10^{-6}\, i$ & $0.09997 +9.33838\times 10^{-6} \, i$ \\ \hline
\end{tabular}
\caption{Quasi-stationary frequencies of states $n=2,m_{l}=-1$ for different sign configurations for $m = 0.1, r_s = 1, P = 0.1, Q = 0.1$ and $a = 0.1$.}
\label{tab:quasi_freqml-1}
\end{table}

\twocolumngrid

It is instructive to compare those results with the present analysis, which is carried out in ingoing Eddington-Finkelstein coordinates extended beyond the outer and inner horizons and therefore provides a novel horizon-regular formulation of the problem. By matching the corresponding energy equations with the quantization condition \eqref{quanfor}, together with the confluent Heun parameters $\alpha_{\pm}$, $\beta_{\pm}$, and $\gamma_{\pm}$ given in \cref{alpha,beta,gamma}, we find that the four modes identified in Ref.~\cite{Bunyaratavej:2024qgk} correspond precisely to the branches associated with $\alpha_{+}$. Specifically, the branches denoted by $(\alpha_{-},\beta_{\pm},\gamma_{\pm})$ in Ref.~\cite{Bunyaratavej:2024qgk} are exactly equivalent to the $(\alpha_{+},\beta_{\mp},\gamma_{\mp})$ branches in the present work.

Despite the use of different coordinate systems and distinct mode classifications, both analyses lead to the same physical conclusion. Quasi-stationary modes with positive energy possess positive imaginary parts, corresponding to states that grow exponentially in time. Such unstable modes are expected to backreact on the geometry and destabilize the region containing closed timelike curves, thereby providing support for the chronology protection conjecture in the non-extremal dyonic Kerr-Sen black hole spacetime. Since the dyonic Kerr-Sen solution represents the most general axisymmetric black hole of the string-inspired Einstein-Maxwell-dilaton-axion theory, this semiclassical result is expected to extend naturally to all simpler rotating black hole solutions.

\section{Conclusions and Discussion}
In this work, we have investigated the quasi-stationary states of a massive scalar field in the background of a dyonic Kerr-Sen black hole. A key element of our analysis is the construction of a horizon-regular coordinate system based on ingoing Eddington-Finkelstein coordinates. This choice provides a smooth extension across the event horizon and allows the ingoing boundary condition to be imposed in a direct and unambiguous manner, in contrast to Boyer-Lindquist coordinates where coordinate singularities obscure the near-horizon behavior.

We formulated the covariant Klein-Gordon equation in this geometry and showed that, despite the change of coordinates, the equation remains separable. The angular sector reduces to the spheroidal harmonic equation, with solutions given by axially symmetric spheroidal harmonics. This confirms that the underlying separability structure of the spacetime is preserved. We then turned to the radial equation and derived an exact analytic solution without resorting to approximation schemes. The quasi-stationary spectrum emerges from imposing the series truncation condition on the radial solution that leads to an exact quantization condition on the frequency. 

The exact solution reveals a rich structure consisting of four distinct branches, corresponding to different parameter choices. We further examined how the spectrum depends on the black hole parameters. Two qualitatively different classes of modes emerge. The branches $(\alpha_+,\,\beta_\pm,\,\gamma_\mp)$ are independent of the rotation and charges, indicating that they are primarily controlled by the scalar mass and the overall gravitational scale. In contrast, the branches $(\alpha_+,\beta_+,\gamma_+)$ and $(\alpha_+,\beta_-,\gamma_-)$ depend explicitly on the rotation parameter $a$ and the charge scale $r_D$, reflecting  sensitivity to the spacetime structure. 

For non-spherical modes, we find that the physical quasi-stationary spectra, where $0< \omega_R<m, \, 
 \mathrm{Re}(\alpha)>0, \, \mathrm{Im}(\alpha)<0 $, exist in the $(+,+,\pm)$ branches for co-rotating modes, while for counter-rotating modes they exist in the $(+,+,-)$ branches. Furthermore, increasing either the electric charge $Q$ or the magnetic charge $P$ produces a symmetric displacement of the spectrum in the complex frequency plane. In particular, $|\operatorname{Re}(\omega)|$ increases, corresponding to higher oscillation frequencies, while $|\operatorname{Im}(\omega)|$ also grows, indicating stronger damping or enhanced instability rates.

For the purely imaginary solutions, we find that two branches correspond to damped non-propagating modes, while the remaining two exhibit unstable behavior. The branches $(\alpha_{+},\beta_{+},\gamma_{-})$ and $(\alpha_{+},\beta_{-},\gamma_{+})$ remain independent of the black hole rotation and charges, whereas $(\alpha_{+},\beta_{+},\gamma_{+})$ and $(\alpha_{+},\beta_{-},\gamma_{-})$ retain an explicit dependence on $a$ and $r_D$. In all cases, the magnitude of the imaginary frequencies increases with the overtone number $n$. Furthermore, these purely imaginary modes possess no oscillatory component and therefore do not propagate through the spacetime. 

We compared the results obtained in this work with those derived in Ref.~\cite{Bunyaratavej:2024qgk}. While the earlier analysis was carried out in Boyer-Lindquist coordinates, the present study employs ingoing Eddington coordinates and therefore offers a novel horizon-regular formulation of the problem. Remarkably, we find that the four physical modes associated with $\alpha_{+}$ in the present work correspond {\bf exactly} to those previously identified in Ref.~\cite{Bunyaratavej:2024qgk}. More precisely, the branches $(\alpha_{+},\beta_{\pm},\gamma_{\pm})$ obtained here are equivalent to the $(\alpha_{-},\beta_{\mp},\gamma_{\mp})$ branches of the earlier work. Hence, despite the different coordinate systems and alternative mode classifications, both analyses lead to the same qualitative physical conclusions.

The common conclusion is that quasi-stationary modes with positive energy possess positive imaginary parts, corresponding to states that grow exponentially in time. Such unstable modes are expected to backreact on the geometry and exponentially deform the spacetime region where closed timelike curves exist, thereby supporting the chronology protection conjecture in the non-extremal dyonic Kerr-Sen black hole spacetime. Moreover, since the dyonic Kerr-Sen solution represents the most generic axisymmetric black hole of the string-inspired Einstein-Maxwell-dilaton-axion theory, this semiclassical evidence is expected to extend naturally to all simpler rotating black hole solutions.

\begin{acknowledgments}
We would like to thank Roman Konoplya for pointing out reference error in our previous manuscript.
\end{acknowledgments}

\appendix

\section{Normal Form} \label{AppendixA}

The normal form of a second-order linear ordinary differential equation is obtained by eliminating the first-derivative term through an appropriate change of dependent variable. This representation is particularly useful for studying qualitative properties of the solutions, such as oscillatory behavior, turning points, and asymptotic structure \cite{NIST}.

We begin with the general linear second-order ordinary differential equation
\begin{equation}
\frac{d^{2}y}{dx^{2}}+p(x)\frac{dy}{dx}+q(x)y=0.
\label{general ODE}
\end{equation}

The first-derivative term can be removed by the standard Liouville transformation
\begin{equation}
y(x)=Y(x)\exp\!\left[-\frac{1}{2}\int p(x)\,dx\right].
\end{equation}
Differentiating, one obtains
\begin{align}
\frac{dy}{dx} &= \exp\!\left[-\frac{1}{2}\int p\,dx\right] \left( \frac{dY}{dx}-\frac{1}{2}pY \right),\\ 
\frac{d^{2}y}{dx^{2}} &= \exp\!\left[-\frac{1}{2}\int p\,dx\right] \left( \frac{d^{2}Y}{dx^{2}} -p\frac{dY}{dx} -\frac{1}{2}\frac{dp}{dx}Y +\frac{1}{4}p^{2}Y \right).
\end{align}

Substituting these expressions into Eq.~\eqref{general ODE}, the terms proportional to $dY/dx$ cancel identically, and the equation reduces to the normal form
\begin{equation}
\frac{d^{2}Y}{dx^{2}}+Q(x)\,Y=0,
\end{equation}
where
\begin{equation}
Q(x)=q(x)-\frac{1}{2}\frac{dp}{dx}-\frac{1}{4}p^{2}(x).
\label{normalform}
\end{equation}

The function $Q(x)$ plays the role of an effective potential and determines the qualitative nature of the solutions. In regions where $Q(x)>0$, the solutions are locally oscillatory, whereas for $Q(x)<0$ they are locally exponential (non-oscillatory). Points satisfying $Q(x)=0$ correspond to turning points separating these two regimes. Under suitable regularity and integrability conditions, stronger global statements regarding the number of zeros of the solutions can also be established \cite{2420}.

\section{The Confluent Heun Equation and Its Solutions}\label{AppendixB}

The confluent Heun equation is a linear second-order ordinary differential equation written in the canonical form \cite{Heun}
\begin{multline}
\frac{d^{2}\psi_H}{dx^{2}}+\left(\alpha+\frac{\beta+1}{x}+\frac{\gamma+1}{x-1}\right)\frac{d\psi_H}{dx}\\+\left(\frac{\mu}{x}+\frac{\nu}{x-1}\right)\psi_H=0,
\end{multline}
where
\begin{align}
\mu&=\frac12\left(\alpha-\beta-\gamma+\alpha\beta-\beta\gamma\right)-\eta,\\ \nu&=\frac12\left(\alpha+\beta+\gamma+\alpha\gamma+\beta\gamma\right)+\delta+\eta.
\end{align}

Its general local solution about $x=0$ is
\begin{multline}
\psi_H(x)=A\,\operatorname{HeunC}(\alpha,\beta,\gamma,\delta,\eta,x)\\+B\,x^{-\beta}\operatorname{HeunC}(\alpha,-\beta,\gamma,\delta,\eta,x),
\label{canonicalheun}
\end{multline}
where $A$ and $B$ are constants.

A necessary condition for the confluent Heun function to truncate to a polynomial of degree $n_r$ is
\begin{equation}
\frac{\delta}{\alpha}+\frac{\beta+\gamma}{2}+1=-n_r,\qquad n_r\in\mathbb{Z}_{\ge 0}.
\label{HeunPol}
\end{equation}

To cast the equation into normal form, we identify
\begin{equation}
p(x)=\alpha+\frac{\beta+1}{x}+\frac{\gamma+1}{x-1},
\qquad
q(x)=\frac{\mu}{x}+\frac{\nu}{x-1},
\end{equation}
and introduce the transformation
\begin{equation}
\psi_H(x) = Y(x)\, e^{-\frac12\alpha x} x^{-\frac12(\beta+1)} (x-1)^{-\frac12(\gamma+1)}.
\end{equation}

Then Eq.~\eqref{normalform} becomes
\begin{multline}
\frac{d^{2}Y}{dx^{2}}+\Bigg[-\frac{\alpha^{2}}{4}+\frac{\frac12-\eta}{x}+\frac{\frac14-\frac{\beta^{2}}{4}}{x^{2}} \\+\frac{-\frac12+\delta+\eta}{x-1}+\frac{\frac14-\frac{\gamma^{2}}{4}}{(x-1)^{2}}\Bigg]Y=0.
\end{multline}

Accordingly, the transformed solution $Y(x)$ is
\begin{multline}
Y(x)=e^{\frac12\alpha x}(x-1)^{\frac12(\gamma+1)}
\Bigg[A\,x^{\frac12(1+\beta)}
\operatorname{HeunC}(\alpha,\beta,\gamma,\delta,\eta,x)
\\+B\,x^{\frac12(1-\beta)} \operatorname{HeunC}(\alpha,-\beta,\gamma,\delta,\eta,x)
\Bigg].
\end{multline}

\bibliography{sn-bibliography}

\end{document}